\documentclass[reprint, amsmath, amssymb, aps, prb, floatfix]{revtex4-2}

\usepackage{graphicx}
\usepackage{dcolumn}
\usepackage{bm}

\usepackage{xcolor}  
\usepackage[colorlinks=true, linkcolor=gray, citecolor=gray, urlcolor=gray]{hyperref}

\usepackage{multirow}

\begin{document}
\bibpunct{}{}{,}{s}{}{}
\preprint{APS/123-QED}

\title{Spin-orbit driven $J_{\mathrm{eff}} = 1/2$ magnetism in a d$^7$ triangular-lattice monolayer cobaltate}

\author{Ritwik Das}
\email{intrd@iacs.res.in}
\author{Soumen Basak}
\author{Mohammad Rezwan Habib}
\author{Indra Dasgupta}
\email{sspid@iacs.res.in}
\affiliation{
 School of Physical Sciences, Indian Association for the Cultivation of Science \\
 2A and 2B Raja S.C. Mullick Road, Jadavpur, Kolkata 700 032, India
}

\date{\today}
             
\begin{abstract}

Recent theoretical and experimental advances have identified cobaltates with a high-spin $d^7$ electronic configuration as promising hosts for spin-orbit entangled $J_{\mathrm{eff}} = 1/2$ magnetism that can support bond-dependent exchange interactions. In two-dimensional triangular lattices, the coexistence of such exchange frustration along with usual geometric frustration gives rise to a rich landscape of competing magnetic phases, establishing monolayer triangular $d^7$ cobaltates as a compelling platform for frustrated magnetism. Here we investigate a representative triangular-lattice monolayer cobaltate CoBr$_2$, where first-principles density functional theory (DFT) calculations reveal a dominant nearest-neighbor $t_{2g}$--$e_g$ hopping channel that enhances the ferromagnetic Kitaev-type exchange interactions. In contrast, the nearest-neighbor Heisenberg term is highly sensitive to a direct $t_{2g}$--$t_{2g}$ hopping path and electronic correlations. The magnetic exchange parameters are evaluated using the hopping amplitudes obtained from DFT calculations within an exact diagonalization framework. We construct the first (J$_1$) and third (J$_3$) nearest neighbor Heisenberg exchange dependent $J_1$--$J_3$ magnetic phase diagram in the physically relevant regime and identify multiple competing ground states, including ferromagnetic, stripy, incommensurate and $120^{\circ}$ antiferromagnetic orders. The Luttinger-Tisza analysis further predicts a Z$_2$ vortex crystal phase, while exact diagonalization reveals a bond-nematic phase promoted by quantum fluctuations. Going beyond the conventional bond-independent XXZ picture typically applied to Co$^{2+}$ systems, our results on monolayer CoBr$_2$ establish d$^7$ cobalt dihalides as a promising platform to explore the interplay of long-range Heisenberg and bond-dependent exchange interactions that can stabilize diverse magnetic ground states on a triangular lattice. 

\end{abstract}

\maketitle

\section{Introduction}

The Kitaev model on a two-dimensional honeycomb lattice~\cite{Kitaev_2006} provides an exactly solvable spin-$1/2$ system hosting quantum spin-liquid ground states with fractionalized Majorana excitations and non-Abelian anyons, stimulating extensive theoretical and experimental efforts toward topological quantum computation and unconventional magnetic and transport phenomena~\cite{Knolle_2019,Takagi_2019}. The microscopic realization of such models was established by the seminal Jackeli--Khaliullin mechanism~\cite{Jackeli_2009}, which demonstrated that bond-dependent Kitaev exchange can emerge in $d^5$ transition-metal (TM) systems with edge-sharing octahedra and 90$^{\circ}$ TM--ligand--TM geometry, stimulating vigorous research on spin-orbit entangled $J_{\mathrm{eff}} = 1/2$ magnetism in both honeycomb and triangular lattices. In real materials, however, along with isotropic Heisenberg exchanges, additional anisotropic exchange interactions inevitably arise from lattice distortions and deviations from the ideal geometry, motivating continued exploration of two-dimensional spin--orbit-coupled frustrated magnets from both theoretical and experimental perspectives~\cite{Winter_2017,Trebst_2022}.

Recently, high-spin $d^7$ cobaltates have emerged as a promising platform for realizing bond-dependent Kitaev exchange arising from the spin--orbit entangled $J_{\mathrm{eff}} = 1/2$ Kramers doublet ground state of the Co$^{2+}$ ion in the Mott insulating regime, despite the relatively weaker spin--orbit coupling (SOC) of 3$d$ cobaltates compared to 5$d$ iridates~\cite{Motome_2018,Khaliullin_2018,Motome_2020}. To date, both theoretical and experimental efforts on $d^7$ cobaltates have primarily focused on honeycomb magnets~\cite{Liu_2020,Lin_2021,TSD_2021,Winter_2022,JGPark_2022,SKPandey_2022,HYKee_2023,HYKee_2024}; however, the Jackeli--Khaliullin mechanism~\cite{Jackeli_2009} identified triangular alongside honeycomb lattices as natural hosts for bond-dependent Kitaev interaction~\cite{Bhattacharyya_2023,Kim_2023}. The triangular-lattice geometry introduces intrinsic magnetic frustration, enriching the phase space of $J_{\mathrm{eff}} = 1/2$ magnetism with competing ground states~\cite{Seiji_1984,Li_2015,Trebst_2015,Brink_2016,Punk_2017,Maksimov_2019,Wang_2021}. Determining the effective spin interactions in such systems is therefore crucial for uncovering the underlying magnetic phases.

\begin{figure*}[ht]
\centering
\includegraphics[width=0.9\textwidth]{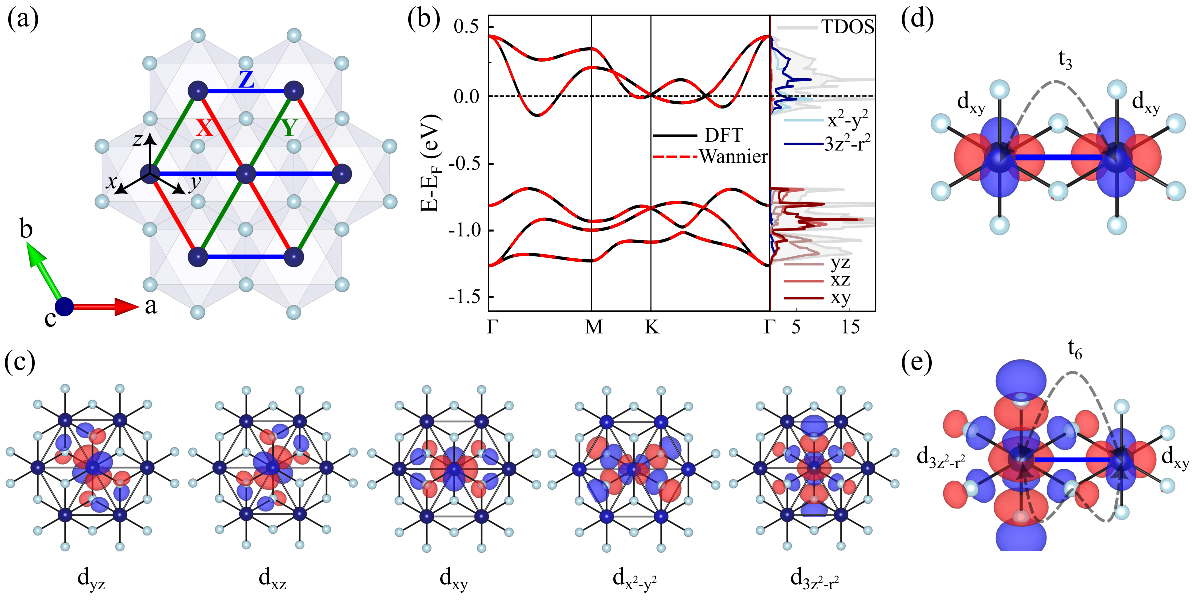}
\caption{\label{DFT}
Crystal structure and non-spin-polarized electronic properties of monolayer CoBr$_2$. 
\textbf{(a)} Crystal structure: Co atoms form a triangular lattice, while Br ligands create edge-sharing octahedra around each Co ion. Local coordinate axes ($x$,$y$,$z$) and the nearest-neighbor bond convention ($X$, $Y$, $Z$) are indicated. 
\textbf{(b)} Non-spin-polarized electronic band structure with Wannier-interpolated bands and partial DOS near $E_F$ obtained from \emph{ab initio} DFT calculations. 
\textbf{(c)} MLWFs for the downfolded Co-$d$ orbitals. 
\textbf{(d)} Direct hopping path $t_3$ between Co-$d_{xy}$ orbitals, primarily contributing to the Heisenberg exchange in $d^7$ cobaltates.
\textbf{(e)} Dominant ligand-assisted hopping path $t_6$, which enhances the Kitaev interaction in monolayer CoBr$_2$.}
\end{figure*}

In contrast to the above proposals, several studies on $d^7$ cobaltates indicate that, in most materials, the nearest-neighbor bond-dependent Kitaev exchange is largely suppressed by competing Heisenberg interactions, leading instead to conventional XXZ-type magnetism arising from the $J_{\mathrm{eff}} = 1/2$ doublets with further-neighbor Heisenberg exchanges~\cite{Lines_1963,Oguchi_1965,TSD_2021,HYKee_2023}. This suppression arises from the relative strength of different hopping channels: indirect ligand-mediated hoppings enhance the Kitaev interaction but typically remain subdominant compared to direct hoppings between the $t_{2g}$ orbitals of the cobalt ions, which instead favor Heisenberg coupling. As a result, realizing strong Kitaev exchange in $d^7$ cobaltates has been considered challenging~\cite{Winter_2022_review,Halloran_2023}. Importantly, our first-principles DFT study of monolayer CoBr$_2$ on the triangular lattice~\cite{CB_strain, CB_DMFT,CB_QAHE} reveals that a dominant ligand-assisted $t_{2g}$--$e_g$ hopping channel strongly favors the Kitaev interaction, establishing monolayer CoBr$_2$ as a promising candidate for exploring $d^7$-based Kitaev magnetism.

In this work, we determine the microscopic magnetic interactions in monolayer CoBr$_2$ using first-principles DFT calculations combined with exact diagonalization (ED) analysis. We examine the evolution of exchange couplings with variations in the relevant microscopic parameters. From the hopping amplitudes obtained through DFT, we find that in addition to the first-nearest-neighbor magnetic exchanges, the third-nearest-neighbor Heisenberg exchange $J_3$ is likely to dominate, while the second-nearest-neighbor coupling $J_2$ remains relatively small. Guided by these insights, we construct the magnetic phase diagram in the $J_1$--$J_3$ parameter space using classical Luttinger-Tisza (LT) and ED methods. Comprehensive studies of the full $J_1$--$K$--$\Gamma$--$\Gamma'$--$J_3$ model for the triangular lattice remain scarce~\citep{Bhattacharyya_2023,Xie_2025}, with only limited attention given to the truncated $J_1$--$K$--$\Gamma$ model~\cite{Rau_2015,Wang_2021}. Building on this context, we map the relevant parameter regime for monolayer CoBr$_2$ and identify multiple competing magnetic orders, including ferromagnetic, stripy, incommensurate (IC), $120^{\circ}$ antiferromagnetic, and Z$_2$ vortex crystal phases. Beyond the classical predictions, ED reveals a bond-nematic phase regime arising from quantum effects and absent in the classical LT framework.

The paper is organized as follows. Section~\ref{section2} presents the crystal and electronic structures along with the hopping parameters obtained from DFT calculations. Section~\ref{section3} introduces the effective spin model and evaluates the corresponding exchange interactions. Section~\ref{section4} analyzes the magnetic ground states and phase diagram using LT and ED methods. Finally, in Section~\ref{section5}, we summarize our findings and conclude.

\section{Crystal structure and non-spin polarized electronic properties}
\label{section2}

\begin{table*}[t]
\centering
\setlength{\tabcolsep}{12pt} 
\renewcommand{\arraystretch}{1.3} 
\caption{CF splittings and hopping parameters (in meV) for monolayer CoBr$_2$ extracted from nonmagnetic DFT calculations.}
\begin{tabular}{c c c c c c c c}
\hline\hline
$\Delta_1$ & $\Delta_2$ & $t_1$ & $t_2$ & $t_3$ & $t_4$ & $t_5$ & $t_6$ \\
\hline
955.6 & 15.7 & 17.4 & 31.7 & -39.1 & -8.9 & 26.7 & 158.7 \\
\hline\hline
\end{tabular}
\label{tab1}
\end{table*}

Bulk CoBr$_2$ crystallizes in the trigonal space group $P\bar{3}m1$ (No.~164) and consists of van der Waals–coupled Br–Co–Br layers, where the magnetism was reported to involve ferromagnetic sheets within each layer coupled antiferromagnetically along the $c$ axis~\cite{CB_bulk}. The corresponding monolayer, derived from the bulk and confirmed to be dynamically stable by first-principles DFT calculations~\cite{CB_strain,CB_DMFT}, is used in the present study. In this monolayer, Co atoms form a two-dimensional triangular lattice with a Co–Co distance of $3.74 \text{\AA}$ and an out-of-plane Co–Br–Co bond angle of $92.6^{\circ}$, while Br ligands form edge-sharing octahedra around Co, as shown in Fig.~\ref{DFT}(a). The local coordinate axes ($x$, $y$, $z$) are defined along the Co–Br bonds, and the three nearest-neighbor Co–Co bonds of the triangular lattice are labeled $X$, $Y$, and $Z$, oriented perpendicular to the corresponding local axes (see Fig.~\ref{DFT}(a)). This convention is followed throughout the paper to evaluate bond-dependent exchange interactions and maintain consistency with the symmetry analysis.

DFT calculations were carried out within the projector augmented-wave (PAW) framework using the \textit{Vienna Ab initio Simulation Package} (VASP) with the generalized gradient approximation for exchange–correlation~\cite{DFT_1, DFT_2, DFT_3}. A plane-wave cutoff of 450 eV and a $\Gamma$-centered $12\times12\times3$ $k$-mesh were employed. Maximally localized Wannier functions (MLWFs) were constructed using \textsc{Wannier90} to extract the nonmagnetic hopping parameters~\cite{Wan_1, Wan_2}.

Fig.~\ref{DFT}(b) shows the non-spin-polarized band structure of monolayer CoBr$_2$, along with the total and partial density of states (DOS). As expected, the Br states are completely occupied (not shown) and the crystal field (CF) offered by Br octahedra splits the Co-d states into distinct t$_{2g}$ and e$_g$ manifolds. These bands primarily arise from antibonding states formed by hybridization between Co-$d$ and Br-$p$ orbitals. The partial DOS for the Co-$d$ orbitals, also shown in Fig.~\ref{DFT}(b), reveals well-separated $t_{2g}$ and $e_g$ manifolds in the local coordinate system defined along the nearly orthogonal Co–Br bonds illustrated in Fig.~\ref{DFT}(a). The resulting $t_{2g}$--$e_g$ CF splitting is $\Delta_1 = 0.96$\,eV. As evident at the $\Gamma$ point in Fig.~\ref{DFT}(b), the $t_{2g}$ manifold further splits into doubly degenerate $e_g^\pi$ levels and a nondegenerate $a_{1g}$ level, with the latter higher in energy, reflecting a trigonal elongation consistent with the $D_{3d}$ site symmetry of Co. 

The Wannierized band structure, retaining only the Co-$d$ bands and downfolding the rest, is in excellent agreement with the non-spin-polarized DFT band structure, as shown in Fig.~\ref{DFT}(b). The maximally localized Wannier functions (MLWFs) for the Co-$d$ orbitals are well aligned with the local coordinate axes (Fig.~\ref{DFT}(c)), revealing a much stronger hybridization of Co-$e_g$ with Br-$p$ orbitals compared to the Co-$t_{2g}$–Br-$p$ coupling. The wannierization further facilitates the construction of a tight-binding (TB) Hamiltonian for the five Co-$d$ orbitals, constrained by the crystal symmetries. 

The corresponding CF matrix takes the form
\begin{align}
\Delta_{CF} =
\begin{pmatrix}
0        & \Delta_2 & \Delta_2 & 0        & 0 \\
\Delta_2 & 0        & \Delta_2 & 0        & 0 \\
\Delta_2 & \Delta_2 & 0        & 0        & 0 \\
0        & 0        & 0        & \Delta_1 & 0 \\
0        & 0        & 0        & 0        & \Delta_1
\end{pmatrix}
\label{eq1}
\end{align}
using the orbital basis ($d_{yz}$, $d_{xz}$, $d_{xy}$, $d_{x^2-y^2}$, $d_{3z^2-r^2}$). Here, $\Delta_2 = 15.7$\,meV denotes the additional splitting of $E_{a_{1g}} - E_{e_g^\pi} = 3\Delta_2 = 47.1$\,meV induced by trigonal elongation. 

Along the nearest-neighbor $Z$ bond (see Fig.~\ref{DFT}(a)), the presence of C$_{2v}$ symmetry constrains the hopping matrix to the form
\begin{align}
T_Z^{ij} =
\begin{pmatrix}
t_1 & t_2 & 0 & 0 & 0 \\
t_2 & t_1 & 0 & 0 & 0 \\
0 & 0 & t_3 & 0 & t_6 \\
0 & 0 & 0 & t_5 & 0 \\
0 & 0 & t_6 & 0 & t_4
\end{pmatrix}
\label{eq2}
\end{align}
between the nearest-neighbor sites $i$ and $j$. The parameters $t_1$–$t_6$ represent the independent symmetry-allowed hoppings extracted from DFT calculations for monolayer CoBr$_2$ (Table~\ref{tab1}). The hopping matrices along the $X$ and $Y$ bonds (Fig.~\ref{DFT}(a)) are generated from $T_Z^{ij}$ by C$_3$-symmetry operations.

\begin{figure*}[ht]
\centering
\includegraphics[width=0.8\textwidth]{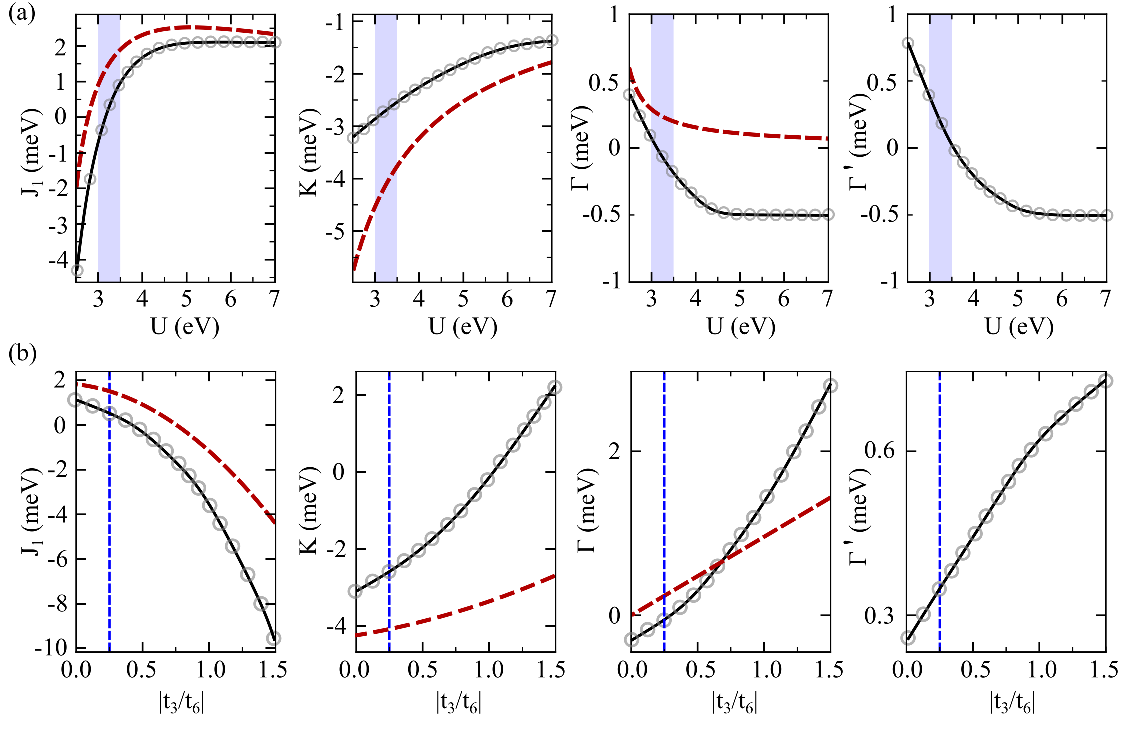}
\caption{\label{Exchange}
Magnetic exchange parameters of monolayer CoBr$_2$ obtained from ED and perturbative calculations. 
\textbf{(a)} Variation of the exchange couplings with the on-site Hubbard interaction $U$ at fixed Hund’s coupling $J_H = 0.7$~eV. 
The physically relevant $U$ range for CoBr$_2$ is shaded in blue. 
\textbf{(b)} Dependence of the exchange couplings on the hopping ratio $|t_3/t_6|$, illustrating the competition between direct and ligand-assisted hopping channels for fixed $U = 3.25$~eV and $J_H = 0.7$~eV. 
Results from ED are shown by circles connected with solid black lines, while perturbative results for $\Delta_2=0$ are indicated by dashed dark-red lines. 
The $|t_3/t_6|$ ratios corresponding to the monolayer CoBr$_2$ is marked by dashed blue vertical line.
}
\end{figure*}

Previous \emph{ab initio} studies on $d^7$ cobaltates have shown that the direct nearest-neighbor hopping $t_3$ between Co-$d_{xy}$ orbitals (Fig.~\ref{DFT}(d)) typically dominates, enhancing the Heisenberg exchange. In contrast, the hopping $t_6$, involving both direct and ligand-mediated processes between $d_{xy}$ and $d_{3z^2-r^2}$ orbitals (Fig.~\ref{DFT}(e)), favors the Kitaev interaction but remains subdominant~\cite{Winter_2022_review,HYKee_2023}, while the other hopping parameters in Eq.~\ref{eq2} are comparatively small. This trend contrasts with the expected suppression of direct hoppings like $t_3$ due to the limited spatial extension of the 3$d$ orbitals, which should instead favor dominant Kitaev exchange through ligand-assisted channels such as $t_6$~\cite{Motome_2018,Khaliullin_2018,HYKee_2024}. 

In monolayer CoBr$_2$, the relatively larger Co–Co bond length suppresses the direct $t_3$ hopping, while the nearly $90^{\circ}$ Co–Br–Co bond angle enhances the ligand-mediated pathways, thereby strengthening $t_6$. Consequently, the $t_6$ channel (Fig.~\ref{DFT}(e)) becomes the largest, expected to favor strong Kitaev exchange, while all other nearest-neighbor hopping amplitudes remain below 40 meV (Table~\ref{tab1}). Moreover, the ligand-mediated $t_2$ hopping between $d_{xz}$ and $d_{yz}$ orbitals along the $Z$ bond is comparable in magnitude to the direct $t_3$ hopping. We observe a similar trend in bulk CoBr$_2$~\cite{Winter_2022_review}, where $t_6$ remains dominant with a value of 177.6\,meV and $t_3$ is smaller at $-90.1$\,meV, though larger than in the monolayer case (Table~\ref{tab1}). The dominant ligand-assisted hopping $t_6$ in monolayer CoBr$_2$ thus provides a microscopic pathway for enhanced Kitaev-type anisotropy. Motivated by these trends, we next analyze how the magnetic exchange interactions evolve with the ratio $t_3/t_6$ to capture the ground state magnetism arising from $J_{\mathrm{eff}} = 1/2$ manifold for monolayer CoBr$_2$.

\section{Microscopic model and exchange interactions}
\label{section3}

In monolayer CoBr$_2$, the Co$^{2+}$ ion adopts a high-spin $S = 3/2$ configuration due to strong Hund’s coupling ($J_H$) relative to the CF splitting. Earlier \emph{ab initio} studies modeled the magnetism of CoBr$_2$ within this $S = 3/2$ framework~\cite{CB_strain,CB_DMFT,Jabar_2021} and reported a ferromagnetic ground state for the monolayer. Recent theoretical works~\cite{Khaliullin_2018,Motome_2018} have shown that beyond the conventional XXZ-type description~\cite{Lines_1963,Oguchi_1965}, the interplay of electronic correlations and SOC within the spin–orbit entangled $J_{\mathrm{eff}} = 1/2$ Kramers doublet derived from the high-spin $d^7$ manifold can give rise to bond-dependent Kitaev exchange interactions. 

The Hilbert space of the Co$^{2+}$ ion corresponds to a high-spin $^4T_{1g}$ term that arises from the combined effects of the on-site Coulomb interaction and the octahedral crystal field. This term splits under SOC into a well-separated $J_{\mathrm{eff}} = 1/2$ Kramers doublet, followed by $3/2$ and $5/2$ multiplets. The $J_{\mathrm{eff}} = 1/2$ doublet accurately describes the ground-state magnetism under trigonal distortion as long as $\Delta_2 \lesssim \lambda/2$~\cite{Lines_1963, Winter_2022_review}. In our case, using $\lambda = 60$~meV~\cite{Lin_2021} yields $\Delta_2 / \lambda \approx 0.26$, confirming its stability and establishing the pseudospin-$1/2$ basis for the effective magnetic model. The single-ion calculation gives $g_{\parallel} = 2.54$ and $g_{\perp} = 5.0$, where $g_{\parallel}$ is defined along the trigonal axis~\cite{Lines_1963}.

To extract the magnetic exchange interactions, we consider the multi-orbital Hamiltonian for a two-site cluster representing the nearest-neighbor Co–Co bond,

\begin{align}
H_{\text{tot}} = H_{\text{CF}} + H_{\text{hop}} + H_{\text{SOC}} + H_U
\label{eq3}
\end{align}

where $H_{\text{CF}}$ and $H_{\text{hop}}$ are obtained from non-spin-polarized DFT calculations, as described in the previous section, and extended to spin space as $\Delta_{\text{CF}} \otimes \mathbf{1}_2$ and $T_{A}^{ij} \otimes \mathbf{1}_2$, where $A$ denotes the nearest-neighbor bond directions $X$, $Y$, or $Z$ shown in Fig.~\ref{DFT}(a). $H_{\text{SOC}}$ represents the atomic SOC, and $H_U$ is the on-site Hubbard–Kanamori interaction~\cite{TSD_2021,HYKee_2023}

\begin{samepage}
\begin{align}
H_{U} &= U \sum_{i,\alpha} n_{i\alpha\uparrow} n_{i\alpha\downarrow}
+ \frac{U'}{2} \sum_{i,\alpha \neq \beta,\, \sigma, \sigma'} n_{i\alpha\sigma} n_{i\beta\sigma'} \nonumber \\
&\quad - \frac{J_{H}}{2} \sum_{i,\alpha \neq \beta,\, \sigma, \sigma'}
c_{i\alpha\sigma}^{\dagger} c_{i\beta\sigma'}^{\dagger} 
c_{i\beta\sigma} c_{i\alpha\sigma'} \nonumber \\
&\quad + \frac{J_{H}}{2} \sum_{i,\alpha \neq \beta,\, \sigma \neq \sigma'}
c_{i\alpha\sigma}^{\dagger} c_{i\alpha\sigma'}^{\dagger} 
c_{i\beta\sigma'} c_{i\beta\sigma} 
\label{eq:HU}
\end{align}
\end{samepage}

where $i$ denotes the site index, $\alpha,\beta$ label the orbitals, and $\sigma,\sigma'$ are spin indices.  
Here, $U$ is the intraorbital Coulomb repulsion, $J_{H}$ is the Hund's coupling, and $U' = U - 2J_{H}$ is the interorbital repulsion. The simplified Hubbard--Kanamori Hamiltonian (Eq.~\ref{eq:HU}), without the three- and four-orbital terms in the Coulomb interaction present in its rotationally symmetric form~\cite{Oles_1984, Spencer_2016}, is found to be adequate as the sizable CF splitting $\Delta_1$ suppresses the effect of these terms in agreement with earlier studies~\cite{Motome_2018, Kee_2022}.

To evaluate the magnetic exchange parameters for monolayer CoBr$_2$, we employ the projection-based ED approach~\cite{Winter_2016,Riedl_2019} on the two-site Hamiltonian in Eq.~(\ref{eq3}). In this method, we take the four lowest energy eigenstates obtained from ED on the two-sites cluster and project them onto the ground-state $J_{\mathrm{eff}}= 1/2$ manifold~\cite{Liu_2020}

\begin{align}
\bigl|\tfrac{1}{2}, \pm\tfrac{1}{2} \bigr\rangle 
&= c_{1} \bigl| \mp 1, \pm\tfrac{3}{2} \bigr\rangle
 + c_{2} \bigl| 0,  \pm\tfrac{1}{2} \bigr\rangle
 + c_{3} \bigl| \pm 1, \mp\tfrac{1}{2} \bigr\rangle
\label{eq:Jeff}
\end{align}

defined in the $\lvert m_{L}, m_{S}\rangle$ basis at each site. The coefficients $c_i$ for $i=1,2,3$ depend on the trigonal CF splitting and SOC strength $\lambda$~\cite{Lines_1963} and adiabatically connected to the pure $J_{\mathrm{eff}}= 1/2$ states as $\Delta_{2} \!\rightarrow\! 0$. This approach goes beyond the conventional perturbative framework by treating the multiplet structure and hybridization effects between states exactly within the two-site cluster. For comparison, we also calculate the exchange parameters using the strong-coupling perturbative expansion~\cite{Motome_2018,Khaliullin_2018,HYKee_2023} in the absence of trigonal distortion (i.e., $\Delta_2 = 0$), providing a useful benchmark for validating the results.

For monolayer CoBr$_2$, the effective nearest-neighbor magnetic interactions can be generalized in the $J_1$–$K$–$\Gamma$–$\Gamma'$ form,
\begin{align}
H_{\text{eff}} =
& \sum_{\langle ij\rangle \in A} 
    J_1\, \vec{S}_i \!\cdot\! \vec{S}_j 
  + K\, S_i^{\gamma} S_j^{\gamma} 
  + \Gamma\, (S_i^{\alpha} S_j^{\beta} + S_i^{\beta} S_j^{\alpha}) \notag\\
& \hspace{2em}
  + \Gamma'\, (S_i^{\alpha} S_j^{\gamma} + S_i^{\gamma} S_j^{\alpha}
             + S_i^{\beta} S_j^{\gamma} + S_i^{\gamma} S_j^{\beta})
\label{Heff}
\end{align}

Here, $\vec{S}_i$ denotes the pseudospin-$\tfrac{1}{2}$ operator which we refer to as spin from now on. $(\alpha,\beta,\gamma)$ label the spin components along the local axes of each Co site. For bonds along the $A=Z$ direction, $(\alpha,\beta,\gamma) = (x,y,z)$, while for $X$ and $Y$ bonds the indices are cyclically permuted to $(y,z,x)$ and $(z,x,y)$, respectively (see Fig.~\ref{DFT}(a)).

The extracted nearest-neighbor exchange parameters for $U = 3.25$~eV and $J_H = 0.7$~eV~\cite{TSD_2021} are summarized in Table~\ref{tab2}, showing a clear dominance of the ferromagnetic Kitaev term ($K < 0$) accompanied by a much smaller antiferromagnetic Heisenberg coupling ($J_1 > 0$). This trend corroborates the microscopic hopping structure discussed in Sec.~\ref{section2}: the strong $t_6$ channel enhances the bond-dependent Kitaev interaction, while the weaker direct $t_3$ hopping between $d_{xy}$ orbitals contributes modestly to the isotropic Heisenberg exchange.

\begin{table}[h]
\centering
\setlength{\tabcolsep}{12pt}
\renewcommand{\arraystretch}{1.3}
\caption{Nearest-neighbor exchange parameters (in meV) for monolayer CoBr$_2$ for $U = 3.25$~eV and $J_H = 0.7$~eV.}
\begin{tabular}{cccc}
\hline\hline
$J_1$ & $K$ & $\Gamma$ & $\Gamma'$ \\
\hline
0.6 & -2.7 & -0.1 & 0.3 \\
\hline\hline
\end{tabular}
\label{tab2}
\end{table}

\begin{figure*}[ht]
\centering
\includegraphics[width=1\textwidth]{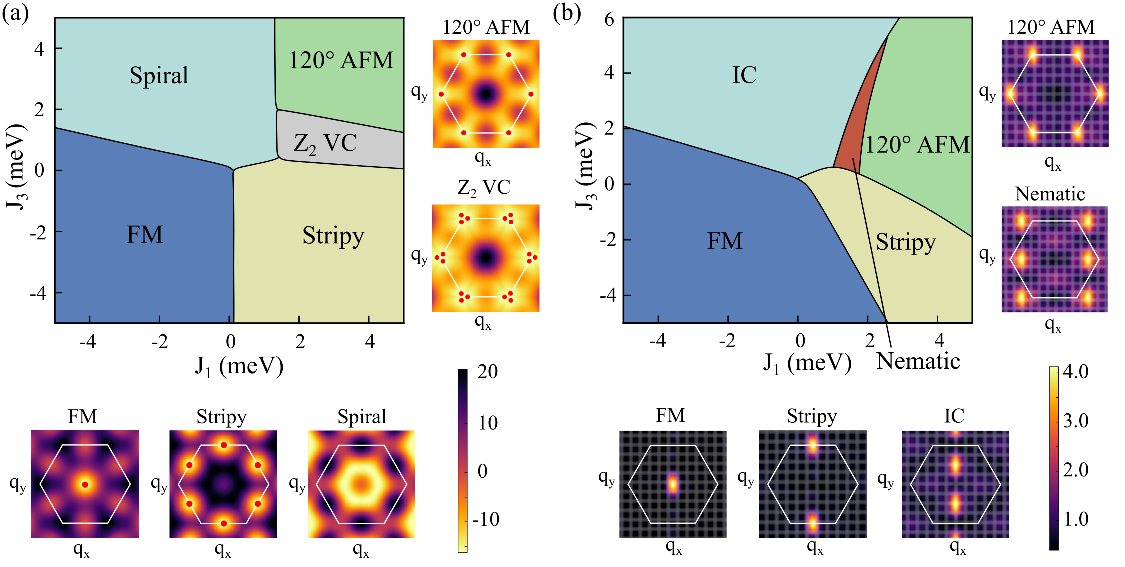}
\caption{\label{phase}
Magnetic phase diagram of monolayer CoBr$_2$. The fixed parameters are $K=-2.7$ meV, $\Gamma=-0.1$ meV and $\Gamma'=0.3$ meV. 
\textbf{(a)}~Semiclassical phase diagram obtained from the LT analysis. 
The reciprocal-space energy distributions for the different phases are shown, with the lowest-energy points (marked by red circles) indicating the ordering wave-vectors corresponding to each phase. Since the wave-vector minima of the spiral phase vary smoothly with the parameters, we show only a representative energy distribution.
\textbf{(b)}~Quantum phase diagram derived from ED calculations on finite triangular-lattice clusters. The static spin structure factors for representative parameter sets within each region are also presented for the cluster shown in Fig.~\ref{ClusterED}(b), illustrating the characteristic magnetic orders. Results for the cluster in Fig.~\ref{ClusterED}(a) are provided in the Supplementary Material~\cite{supplemental}}
\end{figure*}

Figure~\ref{Exchange}(a) shows the evolution of the magnetic exchange parameters with the on-site Coulomb interaction $U$, within the physically relevant range for Co$^{2+}$, at fixed $J_H = 0.7$~eV~\cite{Jabar_2021}, obtained from both perturbative (for $\Delta_2=0$) and ED (for $\Delta_2=15.7$ meV) approaches. The Heisenberg coupling $J_1$ varies sensitively near the preferred value $U = 3.25$~eV and gradually saturates to an antiferromagnetic value ($J_1 \approx 2$~meV) at larger $U$. The Kitaev exchange $K$ remains ferromagnetic ($K < 0$) throughout the entire range, consistent with the perturbative prediction. The off-diagonal coupling $\Gamma$ remains weak, as expected for $3d$ elements, while $\Gamma'$ arises from trigonal distortion. Notably, whereas perturbation theory predicts a consistently antiferromagnetic $\Gamma$, the ED results show a crossover from antiferromagnetic ($\Gamma > 0$) to ferromagnetic ($\Gamma < 0$), reflecting finite trigonal-distortion effects ($\Delta_2$) beyond the idealized limit.

In order to probe the competition between $t_6 (>0)$ responsible for Kitaev exchange K and $t_3 (<0)$ for the nearest neighbor Heisenberg term $J_1$ in monolayer CoBr$_2$, we have examined the dependence of the exchange couplings on the hopping ratio $|t_3/t_6|$, shown in Fig.~\ref{Exchange}(b). As $|t_3/t_6|$ increases, $J_1$ shows a pronounced evolution from antiferromagnetic toward a stronger ferromagnetic character, while $K$ exhibits a crossover from ferromagnetic to antiferromagnetic behavior in the ED results, in contrast to the monotonic ferromagnetic trend predicted by perturbation theory in absence of trigonal distortion. The off-diagonal term $\Gamma$ undergoes a similar ferro–to–antiferromagnetic transition, reflecting the combined effects of trigonal distortion and beyond-perturbative corrections and $\Gamma'$ remains small and antiferromagnetic throughout.

The marked sensitivity of $J_1$ to both $U$ and $|t_3/t_6|$ establishes it as the principal tuning parameter for exploring the magnetic phase diagram. Another important consideration is that the DFT and Wannierization procedures do not account for two-hole virtual exchange processes. Perturbative analyses for high-spin $d^7$ cobaltates show that these processes can significantly enhance the ferromagnetic contribution to the nearest-neighbor Heisenberg term $J_1$~\cite{HYKee_2023,Winter_2022_review} in comparison to other exchange parameters. Related extended treatments of such ligand-mediated contributions within projection-based ED frameworks have also been discussed recently~\cite{Konieczna_2025}. Therefore, the value of $J_1$ listed in Table~\ref{tab2} should be regarded as an estimate subject to such corrections.

Furthermore, beyond the first-nearest-neighbor exchanges, DFT calculations indicate that the largest hopping amplitudes beyond the nearest neighbors correspond to third-nearest-neighbor paths with $|t_{3\mathrm{rd}}^{\mathrm{max}}| \!\sim\! 70$~meV, while second-nearest-neighbor hoppings are suppressed ($|t_{2\mathrm{nd}}^{\mathrm{max}}| \!\sim\! 20$~meV). This trend suggests that the non-negligible third-nearest-neighbor Heisenberg exchange $J_3$, supported also by Slater–Koster estimates~\cite{Winter_2022_review}, may play an essential role in determining the magnetic ground state. Given the broad parameter space and competing interactions, it is therefore instructive to explore the magnetic phase diagram by systematically varying $J_1$ and $J_3$ while keeping $K$, $\Gamma$ and $\Gamma'$ fixed, in order to identify the possible magnetic ground states of monolayer CoBr$_2$. Accordingly, we study the ground state magnetism for the extended Hamiltonian:
\begin{equation}
H = H_{\text{eff}} + \sum_{\langle\langle\langle ij\rangle\rangle\rangle} J_3 \vec{S}_i\cdot \vec{S}_j
\end{equation}

\section{Magnetic ground states and phase diagram}
\label{section4}

For the values of $K$, $\Gamma$, and $\Gamma'$ listed in Table~\ref{tab2}, we map the $J_1$--$J_3$ phase diagram of the spin-$1/2$ triangular lattice using the classical Luttinger-Tisza method~\cite{LT_1946,Litvin_1974,Rau_2014}. Phase boundaries are identified from peaks in the second derivative of the ground-state energy $-\partial^{2}E_{0}/\partial J_{1,3}^{2}$, while the ordering wave-vector $\mathbf{Q}$ is determined by minimizing $E_0$. The resulting phase diagram and corresponding energy distribution in reciprocal space for distinct phases are shown in Fig.~\ref{phase}(a). In addition to ferromagnetic (FM), stripy, and $120^{\circ}$ antiferromagnetic orders, we identify IC spiral and $Z_2$ vortex crystal phases, consistent with previous analyses of the $J$–$K$–$\Gamma$ model on the triangular lattice for $K < 0$ and $\Gamma < 0$~\cite{Rau_2015}.

The FM phase is identified from an energy minimum at the $\Gamma$ point of the Brillouin zone (BZ), while the stripy phase corresponds to minima at the M points. In the IC spiral regime, the LT energy minima deviate from high-symmetry points, indicating a spatially modulated spin configuration. The 120$^{\circ}$ AFM phase is characterized by an energy minimum located at the BZ corners, as shown in Fig.~\ref{phase}(a). In contrast, for the $Z_{2}$ VC phase, the ordering wave-vector is displaced from these high-symmetry points, and this shift occurs differently for the $x$, $y$, and $z$ spin components along their respective nearest-neighbor bond directions $X$, $Y$, and $Z$ (see Fig.~\ref{DFT}(a)). This continuous departure from the BZ corners reflects the instability of the 120$^{\circ}$ AFM state toward a noncoplanar, spatially modulated configuration, giving rise to the topologically nontrivial vortex-crystal texture~\cite{Trebst_2015,Brink_2016}. (See Supplementary Materials~\cite{supplemental} for details.)

Moreover, we observe that the transition between the $Z_2$ VC and 120$^{\circ}$ AFM orders are further influenced by the off-diagonal couplings $\Gamma$ and $\Gamma'$. Specifically, small variations in $\Gamma$ and $\Gamma'$ selectively stabilize either phase: negative $\Gamma'$ favors the $Z_{2}$ VC order, whereas positive $\Gamma'$ promotes the 120$^{\circ}$ AFM state, with $\Gamma$ playing a comparatively minor role in shifting the phase boundaries.

A finite $J_2$ hardly affects the phase diagram. This was confirmed by examining the $J_1$--$J_3$ phase diagram for a physically reasonable range of $J_2 \in [-0.5, 0.5]$~meV as guided from DFT, which produces only minor shifts in the phase boundaries, mainly between the IC spiral and 120$^{\circ}$ orders.

\begin{figure}
\centering
\includegraphics[width=\columnwidth]{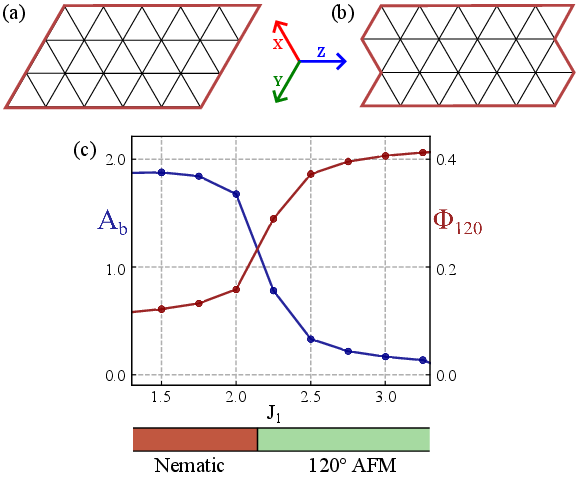}
\caption{\label{ClusterED}
\textbf{(a)--(b)} Finite clusters used in ED calculations. \textbf{(c)} Bond anisotropy $A_{b}$ and the 120$^{\circ}$ AFM order parameter $\Phi_{120}$ for $J_{3}=2.0$~meV as functions of $J_{1}$.}
\end{figure}

To examine the influence of quantum fluctuations, we perform ED calculations~\cite{Sandvik_2010} for the spin-$1/2$ triangular lattice on two distinct $6\times4$ clusters with periodic boundary conditions (see Fig.~\ref{ClusterED}(a) and (b))~\cite{Li_2015, Gong_2017}. The phase boundaries are obtained by combining results from these clusters, as each geometry probes different ordering tendencies and full triangular symmetry cannot be preserved in such finite clusters~\cite{Schmidt_2017}. The magnetic orders are identified from the static spin structure factor, $S(\mathbf{q}) = \langle \vec{S}_{-\mathbf{q}} \cdot \vec{S}_{\mathbf{q}} \rangle$, and the results are shown in Fig.~\ref{phase}(b) together with the ED-derived $J_1$-$J_3$ phase diagram.

Antiferromagnetic $J_1$ and $J_3$ support 120$^{\circ}$ AFM order, which is identified from the peaks of $S(\mathbf{q})$ at the BZ corners (see Fig.~\ref{phase}(b)), a feature very similar to the energy minima obtained by the LT method shown in Fig.~\ref{phase}(a). This is further corroborated by the nearly isotropic nearest-neighbor bond correlations $\langle \vec{S}_i\cdot \vec{S}_{i+A}\rangle\approx |S|^2 cos(2\pi/3) = -0.125$ with $A=X,Y,Z$. The Z$_2$ VC phase identified in the LT method cannot be resolved in the ED analysis due to finite size effects leading to broadening of peaks at BZ corners. A recent experiment on a triangular spin-$1/2$ magnet~\cite{2025_Z2VC_exp} observing a $Z_2$ VC phase reinforces the LT prediction beyond the resolution limits of finite-size ED. Nevertheless, quantum fluctuations weaken perfect coplanarity relative to canonical 120$^{\circ}$ AFM order, as diagnosed by the spin-inertia tensor defined as~\citep{Brink_2016}

\begin{align}
\mathbb{I}^{\alpha\beta}
  = \frac{1}{N^{2}}
    \sum_{i,j}\!
    \left[
      \langle S_{i}^{\alpha} S_{j}^{\beta} \rangle
      - 
      \langle S_{i}^{\alpha} \rangle
      \langle S_{j}^{\beta} \rangle
    \right]
\label{eq:spin-inertia-tensor}
\end{align}

where $\alpha, \beta \in \{x, y, z\}$.  
The eigenvalues of $\mathbb{I}$ distinguish the spin alignment: two vanishing eigenvalue indicates collinear order, one vanishing eigenvalue corresponds to coplanar order, and three comparable eigenvalues signify noncoplanar spin correlations, a characteristic feature associated with vortex-like states such as the Z$_2$ VC.

The sizable ferromagnetic region obtained in our LT and ED analyses is consistent with experimental observations for the bulk system~\cite{CB_bulk} and with earlier \emph{ab initio} predictions for monolayer CoBr$_2$~\cite{CB_strain,CB_DMFT}. The planar anisotropy of this phase is discussed in the Supplementary Material~\cite{supplemental}.

The stripy phase in the phase diagram is identified by peaks at the M point in $S(\mathbf{q})$ for all clusters. The third-neighbor exchange $J_{3}$ is known to promote incommensurate magnetic order~\cite{Hauke_2011,Okubo_2012,Glittum_2021,Bhattacharyya_2023}, which emerge for antiferromagnetic $J_{3}$ and ferromagnetic $J_{1}$. The ED results display consistent features in $S(\mathbf{q})$, characterized by peaks away from the high-symmetry points of the BZ. While classical analysis suggest spiral tendencies in this regime, the precise nature of the quantum ground state can not be definitively resolved within finite-cluster ED. We therefore identify this region as an incommensurate magnetic phase.

ED additionally identifies a bond-nematic regime emerging from quantum fluctuations, absent in the classical LT analysis. To characterize the nematic phase, we compute the bond energy $b_{\alpha}$ along the lattice directions~\cite{Gong_2017} and the bond anisotropy $A_{b}$ defined as 

\begin{align}
 b_{A} = \frac{1}{N_{A}} \sum_{i}\langle \vec{S}_{i} \cdot \vec{S}_{i+A} \rangle, \quad A = \{X, Y, Z\}
\end{align}

\begin{align}
A_{b} = \frac{\lVert b_{A} - \overline{b} \rVert}{\overline{b}}, \quad
\overline{b} = \tfrac{1}{3}\sum_{A} \lvert b_{A} \rvert 
\end{align}
where N$_{A}$ is the number of bonds along the $A$-direction.

A larger value of $A_{b}$, reflecting a dominant $b_A$, signals discrete rotational symmetry breaking. The LT method optimizes only classical spin configurations with factorized single-spin states, and therefore cannot capture orders with primary bond correlations. An enhanced $A_{b}$, together with the absence of such features in the LT analysis, indicates a possible bond-nematic order. The $C_{6}\!\rightarrow\!C_{2}$ symmetry breaking is also visible in $S(\mathbf{q})$ for the nematic order along the preferred direction of the chosen cluster as shown in Fig.~\ref{phase}(b). The IC and bond-nematic phases are distinguished by their $S(\mathbf{q})$ patterns, where the proximity of the Bragg peaks to the Brillouin-zone corners in the nematic phase necessitates its distinction from the $120^{\circ}$ AFM phase. We therefore track their respective order parameters in the phase diagram across the two phases. For the nematic phase we use $A_b$, and for the 120$^{\circ}$ AFM order we use $\Phi_{120}$, defined as the sum of the spin-structure-factor intensities at the BZ corners. The three-sublattice magnetization commonly used to identify the 120$^{\circ}$ AFM order also shows the same trend, supporting this measure. Figure~\ref{ClusterED}(c) illustrates the suppression of 120$^{\circ}$ AFM order along a representative cut and the concomitant breaking of $C_{6}$ rotational symmetry in the bond-nematic regime. 

\section{Conclusion}
\label{section5}

Our combined DFT-derived model Hamiltonian and magnetic ground-state analyses using LT and ED methods establish a framework for understanding spin–orbit entangled magnetism in the triangular-lattice monolayer CoBr$_2$ within the $J_1$–$K$–$\Gamma$–$\Gamma'$–$J_3$ model. Unlike previous experimental and \emph{ab initio} studies that favored effective XXZ models with longer ranged isotropic exchanges~\cite{HYKee_2023,TSD_2021,Halloran_2023}, our results emphasize the crucial role of bond-anisotropic exchanges and provide direct support for the theoretical proposal of bond-dependent Kitaev interactions in $d^7$ cobaltates~\cite{Kim_2023,HYKee_2024, Liu_2021}. This behavior can be attributed to the longer Co–Co distance and the nearly 90$^{\circ}$ Co–Br–Co bond geometry with modest trigonal elongation, which enhance the ligand-assisted hopping channel and generate strong bond-dependent interactions in monolayer CoBr$_2$. Recent experiments on the triangular-lattice cobaltate CoI$_2$~\cite{Kim_2023} also support a magnetic model with bond-anisotropic exchange and third-nearest-neighbor Heisenberg interactions.

We further demonstrate that the $J_1$-$J_3$ magnetic phase diagram obtained using two complementary approaches namely LT and ED method in the physically relevant parameter regime for monolayer CoBr$_2$ hosts diverse set of competing magnetic ground states, including conventional FM, stripy, $120^{\circ}$ AFM and IC phases. In addition the LT method predicts a more exotic Z$_2$ VC phase which has been recently identified in a d$^5$ triangular pseudospin 1/2 magnet~\citep{2025_Z2VC_exp}. While the Z$_2$ VC phase is beyond the resolution of our finite size ED approach, it however uncovers emergence of a bond-nematic region for antiferromagnetic $J_1$ and $J_3$, highlighting the essential role of quantum fluctuations captured in ED methods. The experimental and \emph{ab initio} proposals are well reproduced in our analysis and the previously reported strain-induced FM to AFM transition~\cite{CB_strain} is likewise consistent with the magnetic phase diagram. Our findings provide a basis for understanding the competing bond dependent exchange interactions that govern magnetism in two-dimensional triangular lattice cobaltates and offer a foundation for future exploration of their quantum phases.

\section*{Acknowledgements}
R.D thanks the Council of Scientific and Industrial Research (CSIR), India for research fellowship (File No. 09/080(1171)/2020-EMR-I). S.B acknowledges Science and Engineering Research Board (SERB) India (Project No. CRG/2021/003024). M.R.H acknowledges SERB, India for providing the financial support through National Post Doctoral Fellowship (Project No. PDF/2021/003366). I.D would like to thank Technical Research Center (TRC), Department of Science and Technology (DST) Government of India for support.


%

\clearpage
\onecolumngrid
\bibpunct{}{}{,}{s}{}{}

\title{Supplementary Material for \\
Spin-orbit driven $J_{\mathrm{eff}} = 1/2$ magnetism in a d$^7$ triangular-lattice monolayer cobaltate}

\author{Ritwik Das}
\author{Soumen Basak}
\author{Mohammad Rezwan Habib}
\author{Indra Dasgupta}
\affiliation{
 School of Physical Sciences, Indian Association for the Cultivation of Science \\
 2A and 2B Raja S.C. Mullick Road, Jadavpur, Kolkata 700 032, India
}

\date{\today}
\maketitle
\section{Supplemental Material}
\subsection{Planar anisotropy of the effective spin model}
\noindent
In this section, we re-write the nearest-neighbor effective Hamiltonian presented in Eq.~(6) of the main text as a bond-independent XXZ interaction with bond-dependent anisotropic terms, a representation widely used in the literature~\cite{White_2019, Liu_2021, Winter_2022, Kim_2023}. We then examine the magnetic anisotropy at representative points within the ferromagnetic phase, where the uniform spin configuration provides a clear assessment of the intrinsic anisotropy among the magnetic phases in the $J_1$--$J_3$ phase diagram (Fig.~3 of the main text).
\subsubsection{Transformation to the trigonal basis}
\begin{figure}[h]
    \centering
    \includegraphics[width=0.4\textwidth]{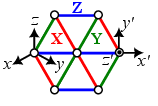}
    \caption{Local octahedral coordinate system $(x,y,z)$ used in the manuscript (see Fig.~1(a)) together with the transformed coordinate system $(x',y',z')$ on the triangular lattice for monolayer CoBr$_2$. The local axes follow the Co--Br bond directions, while the transformed frame is aligned with the trigonal axis to facilitate the analysis of magnetic anisotropy.}
    \label{S1}
\end{figure}

\noindent
In the main text, the effective Hamiltonian is expressed in the local octahedral coordinate system $(x,y,z)$ aligned with the Co–Br bonds (see Fig.~1(a) in the main text). To make the magnetic anisotropy manifest, we transform to a basis $(x',y',z')$ where the $z'$ axis lies along the trigonal direction, i.e., $\hat{z}' = \frac{1}{\sqrt{3}}(\hat{x}+\hat{y}+\hat{z})$, and $x'$ is chosen along the nearest-neighbor $Z$ bond of the triangular lattice~\cite{White_2019, Winter_2022}. In this basis (see Fig.~\ref{S1}), the nearest-neighbor Hamiltonian on a bond $\langle ij\rangle$ can be written as $H_{\mathrm{eff}} = \sum_{\langle ij\rangle} H_{ij}$ with
\begin{align}
\label{eq_1}
H_{ij} &= J_{x'y'}\!\left(S_i^{x'}S_j^{x'} + S_i^{y'}S_j^{y'}\right)
+ J_{z'} S_i^{z'}S_j^{z'} \nonumber\\
&\quad + 2J_{\pm\pm} \!\left[(S_i^{x'}S_j^{x'} - S_i^{y'}S_j^{y'})\cos\phi_{\alpha}
- (S_i^{x'}S_j^{y'} + S_i^{y'}S_j^{x'})\sin\phi_{\alpha}\right] \nonumber\\
&\quad + J_{z'\pm} \!\left[(S_i^{y'}S_j^{z'} + S_i^{z'}S_j^{y'})\cos\phi_{\alpha}
- (S_i^{x'}S_j^{z'} + S_i^{z'}S_j^{x'})\sin\phi_{\alpha}\right]
\end{align}
where $\phi_{\alpha} = (2\pi/3,-2\pi/3,0)$ for nearest-neighbor bonds $X,Y,Z$, respectively. The first two terms in the right-hand side of Eq.~\ref{eq_1} correspond to the bond-independent XXZ interaction, while the remaining terms represent bond-dependent anisotropic couplings.

\noindent
The couplings are related to the parameters $(J_1,K,\Gamma,\Gamma')$ of Eq.~(6) in the main text by~\cite{Liu_2021, Winter_2022}
\begin{align}
J_{x'y'} &= J_1 + \frac{1}{3}(K-\Gamma-2\Gamma') \nonumber\\
J_{z'}   &= J_1 + \frac{1}{3}(K+2\Gamma+4\Gamma') \nonumber \\
J_{\pm\pm} &= -\frac{1}{6}(K+2\Gamma-2\Gamma') \nonumber \\
J_{z'\pm} &= -\frac{\sqrt{2}}{3}(K-\Gamma+\Gamma')
\end{align}

\noindent
Substituting the values of $K$, $\Gamma$, and $\Gamma'$ from Table II of the main text, we obtain
\begin{equation}
\label{eq_2}
J_{z'} = J_1 - 0.6 \ \mathrm{meV}, \qquad
J_{x'y'} = J_1 - 1.1 \ \mathrm{meV},\qquad
J_{\pm\pm} = 0.6 \ \mathrm{meV},\qquad
J_{z'\pm} = 1.1 \ \mathrm{meV}
\end{equation}

\noindent
For the value $J_1 = 0.6$ meV reported in Table II, $J_{z'}$ vanishes. Accounting for a ferromagnetic correction due to two-hole virtual exchange processes ($\delta J$ ranging from $-2$ to $-6$ meV~\cite{Winter_2022}), both $J_{z'}$ and $J_{x'y'}$ become ferromagnetic, with $|J_{z'}| / |J_{x'y'}| < 1$, indicating that spin alignment in the plane perpendicular to the trigonal axis $\hat{z}'$ is energetically favoured. For comparison, experimentally CoI$_2$ yields $|J_{z'}| / |J_{x'y'}| = 0.95$~\cite{Kim_2023}, while for a representative ferromagnetic value $J_1 = -3.0$ meV for monolayer CoBr$_2$, Eq.~\ref{eq_2} gives $|J_{z'}| / |J_{x'y'}| = 0.88$.

\subsubsection{Planar anisotropy in the ferromagnetic phase}

\noindent
We consider a representative point exclusively inside the ferromagnetic phase, $(J_1, J_3) = (-3, -3)$ meV, from both the Luttinger--Tisza (LT) and exact diagonalization (ED) phase diagrams (Fig.~3(a) and (b) in the main text). Diagonalization of the LT interaction matrix (Eq.~\ref{eq_5}) at this point yields eigenvalues $(-21.2, -21.2, -19.7)$ with eigenvectors $(-0.271, -0.532, 0.803)$, $(0.770, -0.620, -0.150)$ and $(0.577, 0.577, 0.577)$ respectively, expressed in the local octahedral coordinate system $(x,y,z)$. Both eigenvectors corresponding to the lowest doubly degenerate eigenvalues are perpendicular to the trigonal direction $\frac{1}{\sqrt{3}}(\hat{x}+\hat{y}+\hat{z})$, thereby indicating easy-plane anisotropy. Other representative points within the ferromagnetic phase yield similar results.

\noindent
To examine the magnetic anisotropy using ED on finite clusters, we use the spin-inertia tensor defined in Eq.~(8) of the main text, which provides a measure of the preferred spin orientations. For the clusters shown in Fig.~4(a) and Fig.~4(b) of the manuscript (denoted here as clusters A and B, respectively), Table~\ref{Tab_S1} lists the eigenvalues and corresponding eigenvectors of the spin-inertia tensor expressed in the local octahedral coordinate system $(x,y,z)$ for the representative ferromagnetic point $(J_1, J_3) = (-3, -3)$ meV. In both clusters, the two eigenvectors associated with the dominant eigenvalues span the plane perpendicular to $\frac{1}{\sqrt{3}}(\hat{x}+\hat{y}+\hat{z})$, which is the eigenvector of the lowest eigenvalue. This demonstrates that the spins preferentially lie in the plane perpendicular to the trigonal axis, consistent with easy-plane anisotropy.

\noindent
As finite clusters do not spontaneously select a particular direction within this plane, the ground state exhibits coplanarity rather than a strictly collinear ferromagnetic order. To verify this further, we introduce a small ferromagnetic pinning field of magnitude $10^{-4}$–$10^{-6}$ meV. The resulting spin-inertia tensor then yields a single dominant eigenvalue with the corresponding eigenvector lying in the plane perpendicular to the trigonal axis, confirming that the collinear ferromagnetic state is confined to this plane.

\begin{table}[h]
\centering
\caption{Eigenvalues and eigenvectors of the spin-inertia tensor $I$ for representative ferromagnetic parameters, expressed in the local octahedral coordinate system $(x,y,z)$. Clusters A and B correspond to Fig.~4(a) and Fig.~4(b) of the main text, respectively.}
\label{Tab_S1}
\begin{tabular}{c c c}
\hline
Cluster & Eigenvalue & Eigenvector \\
\hline
\multirow{3}{*}{A} & 0.505 & $(0.705,\;0.003,\;-0.708)$ \\
                   & 0.494 & $(0.411,\;-0.816,\;0.405)$ \\
                   & 0.000 & $(0.577,\;0.577,\;0.577)$ \\
\hline
\multirow{3}{*}{B} & 0.521 & $(0.408,\;0.408,\;-0.817)$ \\
                   & 0.479 & $(0.707,\;-0.707,\;0.000)$ \\
                   & 0.000 & $(-0.578,\;-0.578,\;-0.577)$ \\
\hline
\end{tabular}
\end{table}

\subsection{Luttinger-Tisza and $Z_2$ VC phase}
\noindent
The LT minimization method~\cite{LT_1946} involves replacing the local spin-length constraints, $\textbf{S}_{i}^{2}=S^2$, $\forall$ i (also known as the strong constraint), with a single, global constraint $\sum_{i}^{}\textbf{S}_{i}^{2}=NS^2$ (weak constraint), where $N$ is the number of spin sites. The classical spin Hamiltonian is then minimized in momentum space under this weak constraint. If the spins corresponding to the energy minima also satisfy the strong constraint, then it is the true classical ground state.\\
\noindent
Numerically one needs to minimize the following energy functional
\begin{equation}
    \varepsilon/N = \sum_q \textbf{S}_{-q}\cdot \tilde{H}(q)\cdot \textbf{S}_{q} - \lambda_0 \left(\sum_q \textbf{S}_{-q}\cdot \textbf{S}_{q} - 1\right)
\end{equation}
\noindent
where $\textbf{S}_{i} = \sum_{q}e^{iq\cdot r_i} \textbf{S}_{q}$ and $\tilde{H}(q)$ a 3 $\times$ 3 matrix, is the Fourier transform of Eq.~(7) in the manuscript. $\lambda_{0}$ is the Lagrange multiplier representing the global constraint. $\partial\varepsilon / \partial \textbf{S}_{-q} = 0$ leads to the eigenvalue equation
\begin{equation}
\label{eq_5}
    \tilde{H}(q) \textbf{S}_{q} = \lambda_{0} \textbf{S}_{q}
\end{equation}
Due to the bond-anisotropic nature of the Hamiltonian, in the $Z_{2}$ VC phase (Fig.~3(a) of the manuscript), the 3 eigen-modes are minimized at 3 distinct incommensurate wave-vectors. The eigen-modes cannot be combined to satisfy the strong constraint on every site. Consequently, the LT method fails to provide an exact classical ground state in this region, and the LT energy serves as a lower bound~\cite{Brink_2016}.\\

\noindent
Despite the violation of the local constraint, the wave-vectors that minimize the LT energy functional in the $Z_{2}$ VC phase have been shown to correspond to the primary Fourier modes of the non-coplanar ground state reported in Monte Carlo simulations for the Heisenberg-Kitaev model~\cite{Brink_2016}. As shown in Fig.~3(a) of the manuscript, the LT energy minima are displaced from the BZ corner points, and are given by $\textbf{q}_{\alpha} = \textbf{K} - t \textbf{a}_{\alpha}$, where $\textbf{a}_{\alpha}$, $\alpha=X,Y,Z$ are the lattice vectors~\cite{Trebst_2015} (see Fig.~4(c) in the manuscript). Thus, while the LT method does not yield an exact ground state, it captures the leading instability that drives the system into the $Z_2$ VC phase.

\subsection{Additional $S(q)$ plots}
\begin{figure}[h]
    \centering
    \includegraphics[width=1\textwidth]{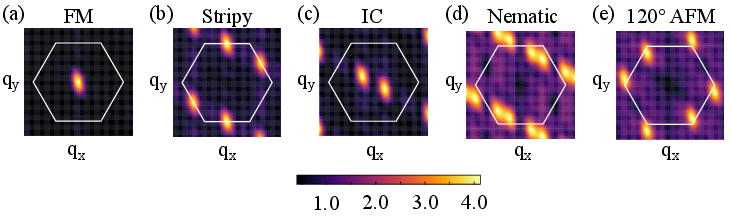}
    \caption{Static spin structure factor $S(q)$ in \textbf{(a)} FM, \textbf{(b)} Stripy, \textbf{(c)} IC, \textbf{(d)} Bond-nematic and \textbf{(e)} 120$^\circ$ AFM phase for the cluster shown in Fig.~4(a) of the manuscript}
    \label{S2}
\end{figure}

\noindent
The presence of two pairs of peaks at different $M$ points in the stripy phase (Fig.~\ref{S2}(b)) arises because the finite cluster permits two possible orientations of the stripy order, and the ground state forms a superposition of these configurations. The slight displacement of the peaks from the corners of the BZ in the 120$^\circ$ phase (Fig.~\ref{S2}(e)) reflects the fact that the cluster is not perfectly commensurate with the corresponding spin configuration. 

\noindent
For the bond-nematic phase (see Fig.~\ref{S2}(d) above and Fig.~3(b) of the manuscript), the two non-equivalent 24-site clusters (Fig.~4(a) and (b) of the manuscript respectively), select different $C_2$ axes. While the orientation of the cluster geometry determines which axis is accessible within a given finite system, the signature of rotational symmetry breaking and suppression of 120$^{\circ}$ AFM order is consistent across both clusters in the same parameter region.

\end{document}